# Synthesis of Graphene on Gold

Tuba Oznuluer[1], Ercag Pince[2], Emre O. Polat[2], Osman Balci[2], Omer Salihoglu[2], Coskun Kocabas[2,*]

[1] *Ataturk University, Department of Chemistry, Erzurum, Turkey*

[1] *Advanced Research Laboratories, Department of Physics, Bilkent University, 06800, Ankara, Turkey*

**ABSTRACT**: Here we report chemical vapor deposition of graphene on gold surface at ambient pressure. We studied effects of the growth temperature, pressure and cooling process on the grown graphene layers. The Raman spectroscopy of the samples reveals the essential properties of the graphene grown on gold surface. In order to characterize the electrical properties of the grown graphene layers, we have transferred them on insulating substrates and fabricated field effect transistors. Owing to distinctive properties of gold, the ability to grow graphene layers on gold surface could open new applications of graphene in electrochemistry and spectroscopy.

*Corresponding author: ckocabas@fen.bilkent.edu.tr*

Graphene, the two-dimensional (2D) crystal of sp2 hybridized carbon, provides unique electronic[1] and mechanical properties[2]. Recent progress on synthesis of graphene on large area substrates[3] fosters new applications of the 2D crystal. High speed electronics[4] and macroelectronics[5] are two examples of these emerging applications. Catalytic decomposition of hydrocarbons on transition metals provides a more promising method for large area growth of high quality graphene layers. Various methods such as chemical vapor deposition (CVD)[3], surface segregation[6], solid carbon source[7] and ion implantation[8] have been used to synthesis of graphene layers on metal surfaces. Nickel[9] and copper[10] are the most commonly used metal substrates for the growth. Other metals such as ruthenium, iridium, platinum, palladium, and cobalt have also been used as a substrate for chemical vapor deposition of graphene[10-13]. Ability to grow graphene on different metals provides opportunities for new applications and more inside to understand the growth mechanism. In this Letter we show a method to directly grow graphene layers on gold surface in various forms such as thin films, foils and wires. The Raman spectroscopy reveals the essential features of the grown graphene layers. In order to characterize the electrical properties of the grown graphene layers, we have transferred them on insulating substrates and fabricated back gated field effect transistors (FET) that use the graphene as an effective semiconductor.

Formation of graphene on metal surfaces has been known for 40 years[13]. Single crystal Ru and Pt are the earliest metals on which epitaxial graphene due to the segregation of carbon impurities has been observed[13]. Recent studies have demonstrated chemical vapor deposition of a single layer graphene on large number of metal substrates. Gold surface, however, has not been used for graphene growth. Previously adsorption of graphene-like polycyclic hydrocarbon across gold step edges has been reported[14]. We believe that the synthesis of large area graphene on gold surface could motivate new applications especially in electrochemistry[15,16] and spectroscopy[17]. It is known that gold surface has no catalytic activity. Gold nanoparticles, however, show some diameter dependent catalytic activity for oxidation of carbon monoxide and alcohols[18]. It is found that the smaller particles are more active for oxidation. Furthermore growing single-walled (SWNT) and multi-walled carbon nanotubes (MWNT) over gold nanoparticles also shows catalytic activity of the gold nanoparticles. Bhaviripudi et al.[19] grew SWNTs on $SiO_2$ substrates using Au nanoparticles. More recently Yuan et al.[20] showed the growth of well aligned SWNT using Au particles. This catalytic activity could be due to the solubility of carbon in gold clusters[21]. The mechanism of catalytic activity of Au particles during CVD process still remains unclear. We believe that the role of gold nanoparticle during the CVD process could provide more inside to understand the growth mechanism of graphene layers on gold surface.

Figure 1 shows the schematic representation of the steps of the preparation of the gold foils, deposition and transfer process of the graphene layers on dielectric substrates. We used 25 µm thick gold foils obtained by pressing high purity gold plates (99.99% from Vakıf Bank). In order to remove impurities and reconstruct the single crystalline surface, the gold foils were annealed with a hydrogen flame for 20 minutes before use. Owing to the fast heating and cooling rates, hydrogen flame annealing provides more complete crystalization of the gold foil than the furnace annealing[22]. The result of this treatment was a polycrystalline gold surface partially (111) oriented, with a roughness lower than a few nanometers. After the annealing step, the gold foil is placed in quartz chamber and the chamber is flushed with Ar gas for 5 min. The foil is heated up to $975^0$ C under Ar and $H_2$ flow (240 sccm and 8 sccm, respectively). Methane gas with a rate of flow of 10 sccm is sent to the chamber for 10 min. After stopping the methane flow, we cooled the chamber with a natural cooling rates around 10 C/sec. We have also developed a transfer printing method to the transfer the graphene from the gold foil to insulating dielectric substrates. After the deposition, an elastomeric stamp PDMS (Polydimethylsiloxane) is applied on the graphene coated gold foil. The gold layer is etched by diluted gold etchant (type TFA, Transene Company Inc.). After complete etching of the gold foil, the graphene layer on PDMS is applied on a 100 nm $SiO_2$ coated silicon wafers. Peeling the PDMS releases the graphene on the dielectric surface.

We grew graphene layers on gold surface in various forms such as thin films, foils and wires. After the growth, we inspect the grown graphene layers using optical microscope and scanning electron microscope. Figure 2(a) shows scanning electron micrograph of the graphene on gold foils, respectively. The inset in Figure 2(a) shows the optical micrograph of the graphene on the gold foil. The grain boundaries and wrinkles on the polycrystalline gold surface are clearly seen.

Raman spectroscopy provides clear finger prints of graphene layers. Raman spectrum of the samples was measured using a confocal microraman system (Horiba Jobin Yvon) in back-scattering geometry. A 532 nm diode laser is used as an excitation source and the Raman signal is collected by a cooled CCD camera. Figure 2(b) and 2(c) shows the Raman spectrum of the graphene as grown on gold and after the transferring on $SiO_2$ wafer. The expected Raman peaks of D, G, D`, and 2D can be seen on the graph. The position of D, G, D`, and 2D peaks are 1371, 1600, 1640, and 2743 $cm^{-1}$ on gold and 1339, 1584, 1620 and 2676 $cm^{-1}$ on $SiO_2$ substrate. There is a significant red shift after the transfer process, likely due to release of compressive strain on the graphene film. The inset in figure 2(c) shows the zoomed 2D peak and the Lorentzian fit. The fit has a symmetric Lorentzian shape with a width of 37 $cm^{-1}$. The shape and the width of the 2D peak is a good indication of single layer graphene or noninteracting a few layers of graphene. The D-band Raman signal of graphene on gold is relatively more intense than the graphene grown on copper. This intense D band is likely due to the large lattice mismatch between gold and graphene. The lattice constant of bulk graphite and hexagonally closed-packed gold surface are 2.46 Å and 2.88 Å respectively.

In order to find the optimum growth conditions we performed Raman spectroscopy on samples grown at temperatures between $850^0$C and $1050^0$C (See Figure 2(d)). At low temperatures around $850^0$C, we do not observe a significant 2D peak. On the other hand at very high temperatures there is a significant D and G peaks which indicates formation of multilayer defective graphene layers. For the samples grown at temperatures between $850^0$C and $1000^0$C, the Raman spectra look relatively the same. We have also studied the effect of the cooling process on the grown graphene layers. However we do not observe any significant change in the Raman spectra for cooling rates between 10 C/sec to 0.5 C/sec.

Figure 3(a) and 3(b) shows micro-Raman mapping images of the graphene layers on $SiO_2$. The micro-Raman mapping images are measured using confocal microraman system with a spatial resolution of 300nm. Figure 3(a) shows the 2D map of the Lorentzian width of 2D peak obtained by scanning the sample and Lorentzian curve fitting for each point. The intensity ratio mapping images of 2D and G peaks is shown in Fig.3(b). The histogram of the Lorentzian width and $I_{2D}/I_G$ peaks are given in Fig. 3(c) and 3(d), respectively. The average width and the intensity ratio is around 40 cm$^{-1}$ and 1.5, respectively. The distributions of the histograms further support high percentage of single layer graphene.

To further characterize the grown graphene layers, we have fabricated field effect transistors in thin film geometry. Source and drain electrodes are fabricated on the transferred graphene layer using a standard UV photolithography. The schematic representation of the graphene FET is shown in Figure 4(a). The devices are isolated from each other by $O_2$ plasma etching of the graphene layers between the devices. A highly doped silicon substrate functions as a global gate electrode and 100 nm thick $SiO_2$ forms the gate dielectric. A semiconductor parameter analyzer (HP 4142B) is used to measure the electrical characteristics of the fabricated transistors. The 2D Raman intensity map of the graphene layer used for the device is shown in Figure 4(b). Figure 4(c) shows the transfer characteristics of a device with a channel length of 8 µm and channel width of 100 µm at a drain bias of 1 V. The Dirac point of the device is around -20 V which is likely because of the trapped charges on the dielectric layer or unintentional doping during the fabrication process. The on-off ratio of the device is around 2. The calculated field effect mobility of the device is around 20 cm$^2$/Vs. The output characteristics are given in Figure 4(d). The modulation of the channel conductivity is clearly seen from the graph.

The Raman spectra together with the transport measurements provide solid evidence that gold surface can be used as a substrate for graphene. Graphene growth on metal substrates shows different growth mechanism depending on the solubility of carbon in the metal. Two different growth mechanisms have been proposed for nickel and copper substrate. Very recently, using sequentially introduced isotopic carbon, Li et al.[23] demonstrated that the growth mechanism on nickel is based on diffusion and precipitation, however, the growth mechanism on copper is based on surface adsorption. Maximum solubility values of carbon in nickel, copper and gold are 2.7, 0.04 and 0.06 % respectively[24]. The solubility in gold is slightly more than copper and much less than nickel. Based on these solubility values and the observed minor effect of cooling rates, we speculate that the growth mechanism of graphene on gold surface could be similar with copper.

In summary we have reported the synthesis of graphene layers on gold surface using chemical vapor deposition technique. The Raman spectra of the samples reveal the essential feature of the graphene grown on gold. The back-gated field effect transistors that use the transferred graphene layers as an effective semiconductor are fabricated and characterized. Further surface characterization experiments are needed to understand the growth mechanism and the nature of the interaction between the gold and graphene. We believe that, gold surface coated with graphene layer could provide a unique configuration for various new applications ranging from surface plasmon resonance sensors to electrochemical analysis.

**Acknowledgement :** This work was supported by the Scientific and Technological Research Council of Turkey (TUBITAK) grant no. 109T259, and Marie Curie International Reintegration Grant (IRG) grant no 256458 . Raman measurements were performed in part at UNAM-Institute of Materials Science and Nanotechnology.

**References:**

1. K. S. Novoselov, A. K. Geim, S. V. Morozov, D. Jiang, Y. Zhang, S. V. Dubonos, I. V. Grigorieva, and A. A. Firsov, Science **306,** 666-669 (2004).
2. A. H. Castro Neto, F. Guinea, N. M. R. Peres, K. S. Novoselov, and A. K. Geim, Reviews of Modern Physics **81,** 109-162 (2009).
3. X. S. Li, W. W. Cai, J. H. An, S. Kim, J. Nah, D. X. Yang, R. Piner, A. Velamakanni, I. Jung, E. Tutuc, S. K. Banerjee, L. Colombo, and R. S. Ruoff, Science **324,** 1312-1314 (2009).
4. Y. M. Lin, C. Dimitrakopoulos, K. A. Jenkins, D. B. Farmer, H. Y. Chiu, A. Grill, and P. Avouris, Science **327,** 662-662 (2010).
5. J. A. Rogers, Nature Nanotechnology **3,** 254-255 (2008).
6. N. Liu, L. Fu, B. Dai, K. Yan, X. Liu, R. Zhao, Y. Zhang, and Z. Liu, Nano Letters **11,** 297-303 (2010).
7. Z. Sun, Z. Yan, J. Yao, E. Beitler, Y. Zhu, and J. M. Tour, Nature **468,** 549-552 (2010).
8. S. Garaj, W. Hubbard, and J. A. Golovchenko, Applied Physics Letters **97,** 183103 (2010).
9. A. Reina, X. Jia, J. Ho, D. Nezich, H. Son, V. Bulovic, M. S. Dresselhaus, and J. Kong, Nano Letters **9,** 30-35 (2008).
10. L. Gao, J. R. Guest, and N. P. Guisinger, Nano Letters **10,** 3512-3516 (2010).
11. J. Coraux, A. T. N'Diaye, M. Engler, C. Busse, D. Wall, N. Buckanie, F. Heringdorf, R. van Gastei, B. Poelsema, and T. Michely, New Journal of Physics **11,** 023006 (2009).
12. E. Sutter, P. Albrecht, and P. Sutter, Applied Physics Letters **95,** 133109 (2009).
13. J. Wintterlin and M. L. Bocquet, Surface Science **603,** 1841-1852 (2009).
14. M. Treier, P. Ruffieux, R. Schillinger, T. Greber, K. Mullen, and R. Fasel, Surface Science **602,** L84-L88 (2008).
15. C. S. Shan, H. F. Yang, J. F. Song, D. X. Han, A. Ivaska, and L. Niu, Analytical Chemistry **81,** 2378-2382 (2009).
16. Y. Y. Shao, J. Wang, H. Wu, J. Liu, I. A. Aksay, and Y. H. Lin, Electroanalysis **22,** 1027-1036 (2010).
17. L. Wu, H. S. Chu, W. S. Koh, and E. P. Li, Optics Express **18,** 14395-14400 (2010).
18. H. Kanzow and A. Ding, Physical Review B **60,** 11180-11186 (1999).
19. S. Bhaviripudi, E. Mile, S. A. Steiner, A. T. Zare, M. S. Dresselhaus, A. M. Belcher, and J. Kong, Journal of the American Chemical Society **129,** 1516-+ (2007).
20. D. N. Yuan, L. Ding, H. B. Chu, Y. Y. Feng, T. P. McNicholas, and J. Liu, Nano Letters **8,** 2576-2579 (2008).
21. N. Yoshihara, H. Ago, and M. Tsuji, Japanese Journal of Applied Physics **47,** 1944-1948 (2008).
22. V. L. De Los Santos, D. Lee, J. Seo, F. L. Leon, D. A. Bustamante, S. Suzuki, Y. Majima, T. Mitrelias, A. Ionescu, and C. H. W. Barnes, Surface Science **603,** 2978-2985 (2009).
23. X. Li, W. Cai, L. Colombo, and R. S. Ruoff, Nano Letters **9,** 4268-4272 (2009).
24. H. Okamoto and T. Massalski, Journal of Phase Equilibria **5,** 378-379 (1984).

**Figure Captions**:

Figure 1: Schematic representation of the steps of deposition and transfer process of graphene growth on gold film.

Figure 2: (a) Scanning electron micrograph of the graphene coated gold foils. The inset shows the optical micrograph of the same area. (b) Raman spectra of the graphene as grown on gold surface. (c) Raman spectra of the graphene on $SiO_2$ substrate. The inset shows the zoomed spectrum of the 2D peak and Lorentzian fit. The width of the Lorentzian is around 37 cm$^{-1}$. (d) Raman spectra of the graphene samples grown at a range of temperatures between 850$^0$C and 1050$^0$ C.

Figure 3: The micro-Raman mapping images of (a) the Lorentzian width of the 2D peak and (b) intensity ratio ($I_{2D}/I_G$) of 2D and G peaks of the graphene on $SiO_2$ layer. The Lorentzian widths are obtained by fitting the spectrum for each point. The scale bars are 5 µm. Histogram of (c) the Lorentzian width of the 2D peak and (d) the intensity ratio ($I_{2D}/I_G$).

Figure 4 (a) Schematic representation of the back-gated graphene field effect transistors. (b) Raman intensity map of 2D peak from the graphene layer bridging the source and drain electrodes. (c) Transfer and (d) output characteristics the transistor with a channel width of 100 µm and a channel length of 8 µm.

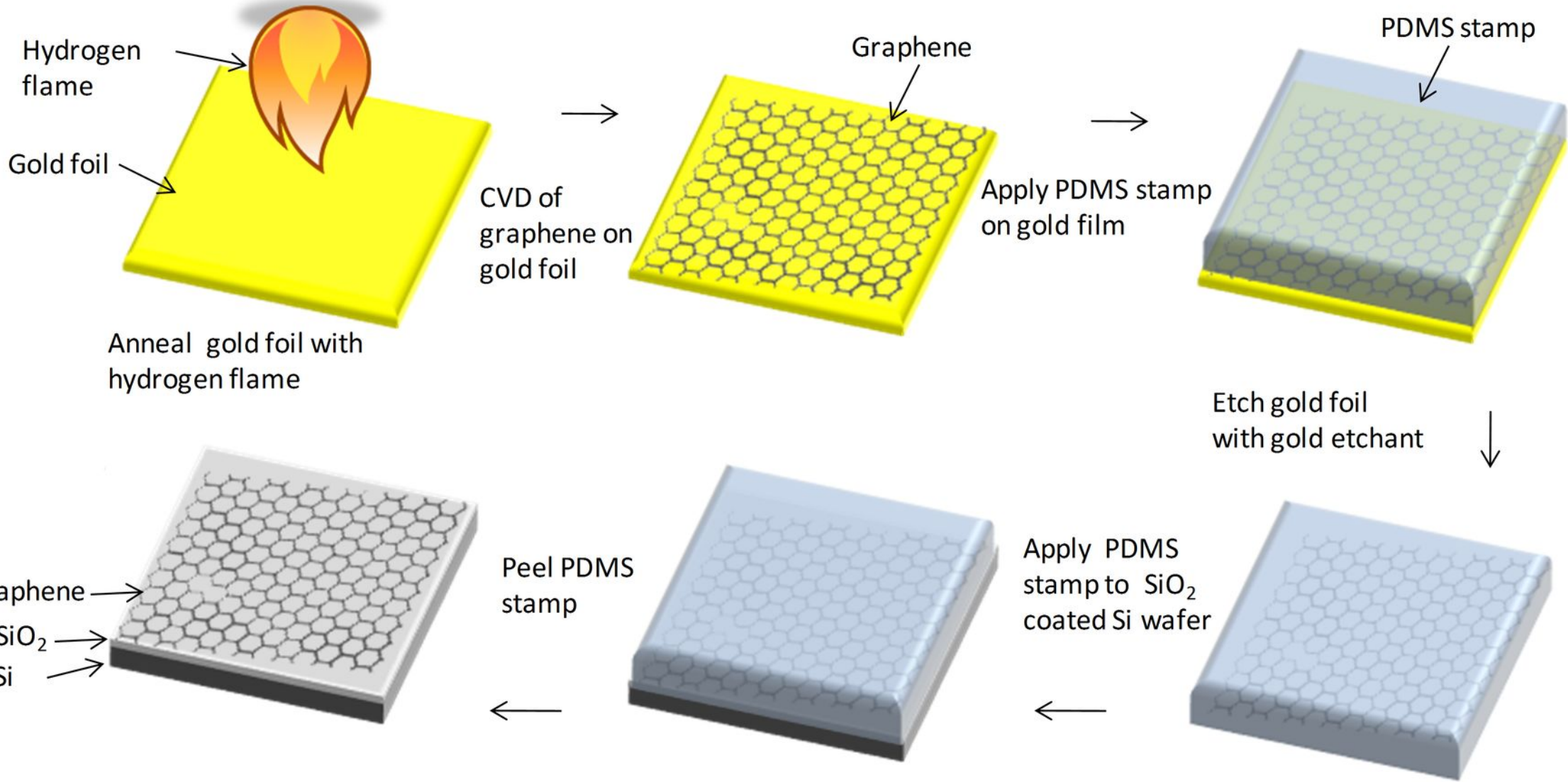

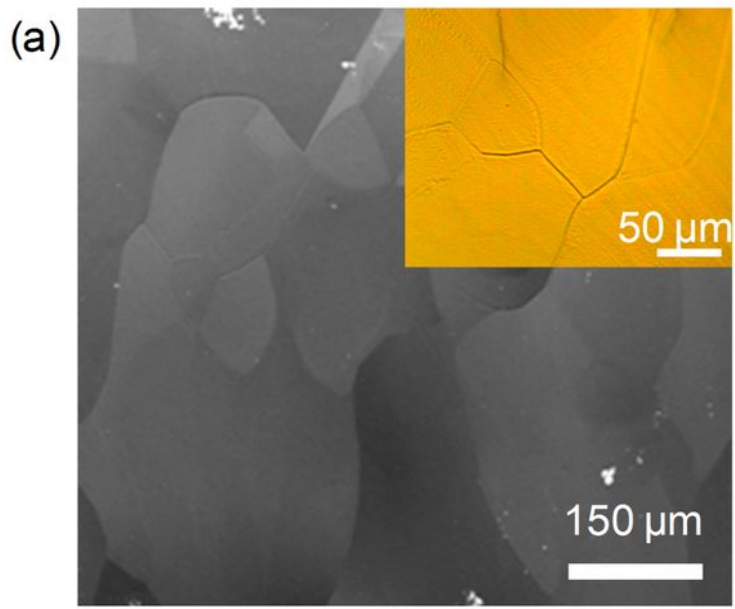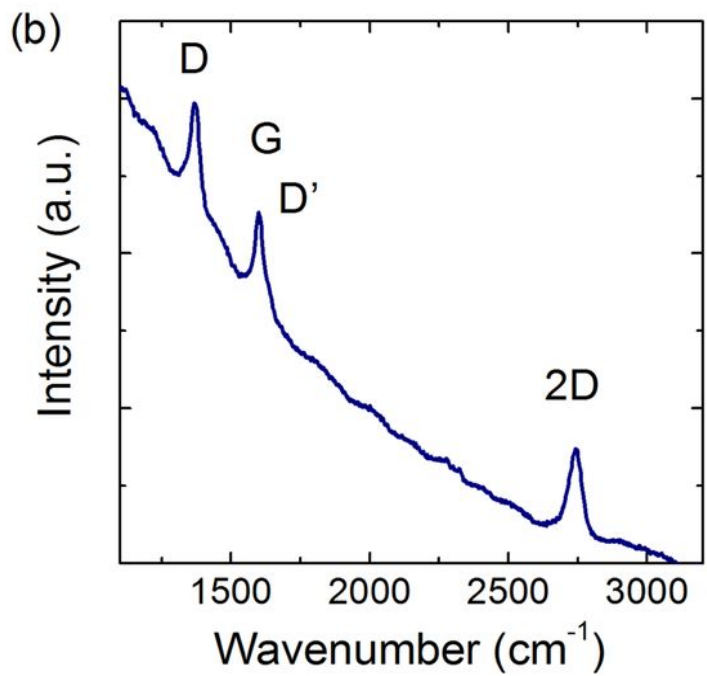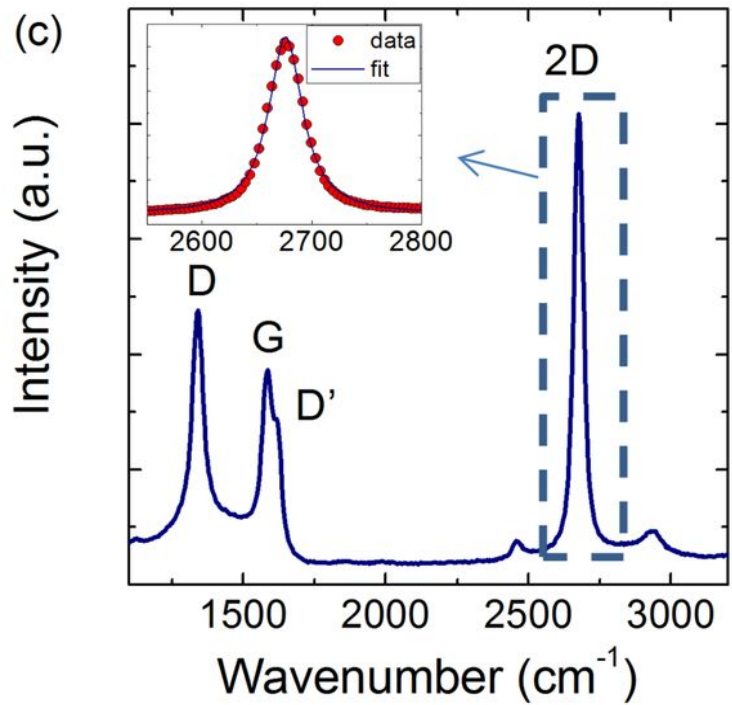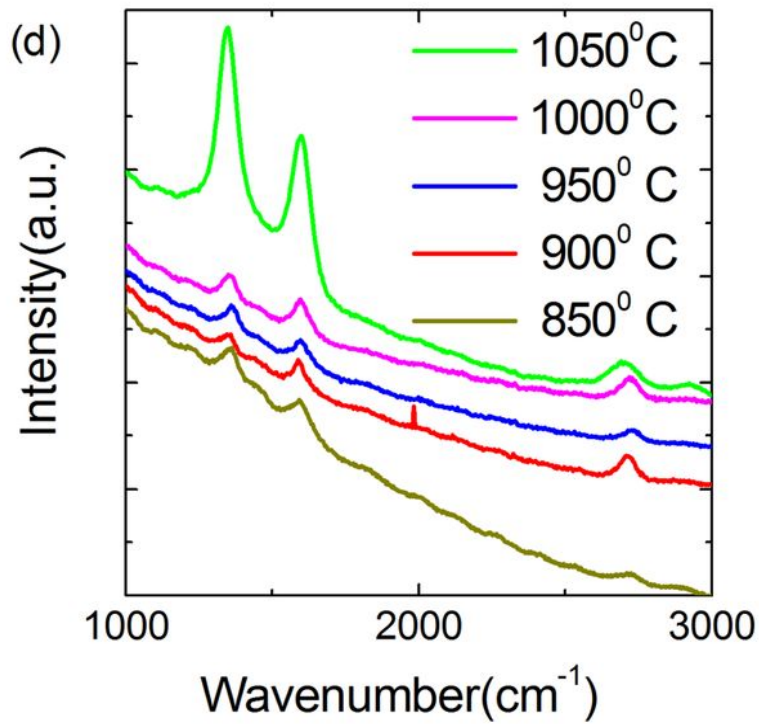

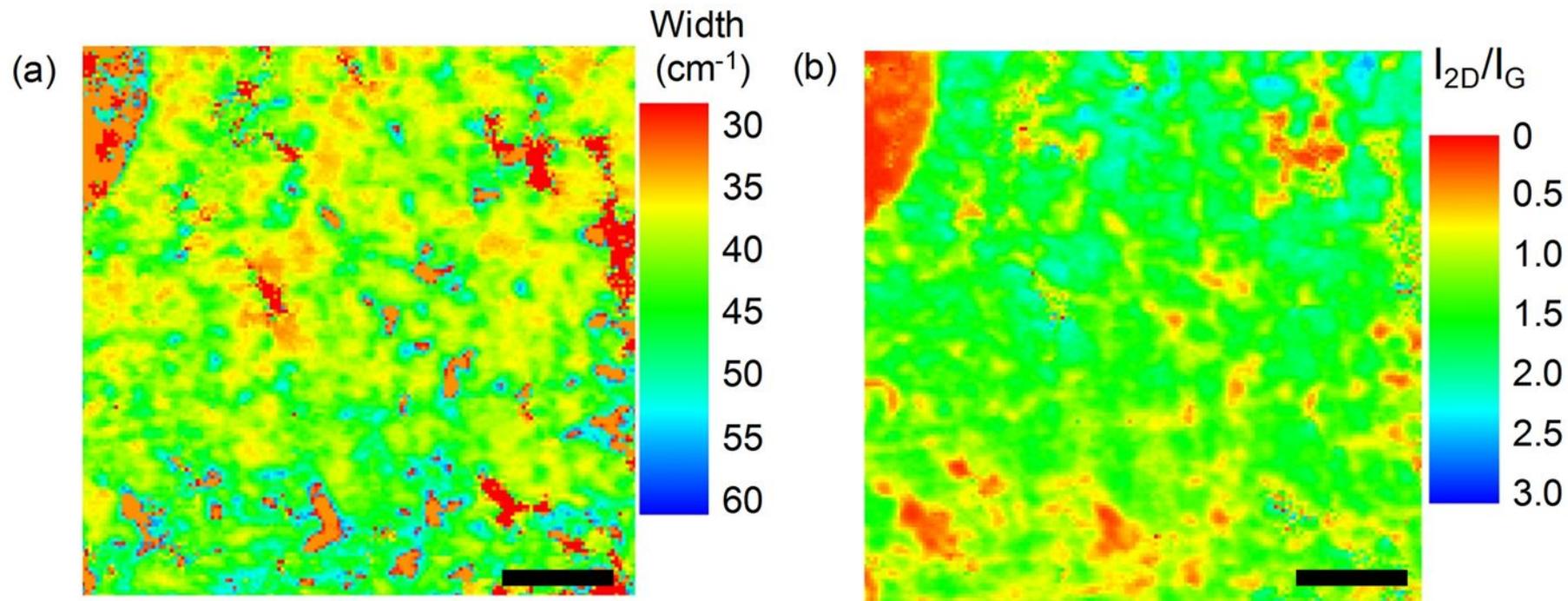

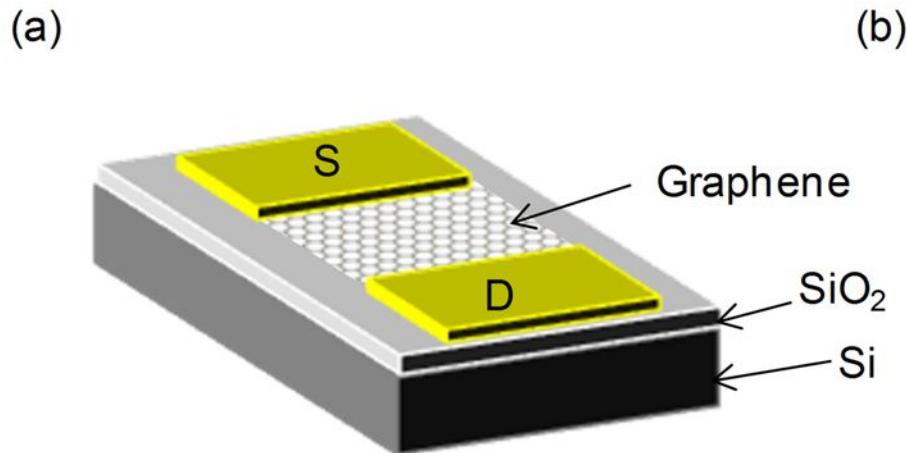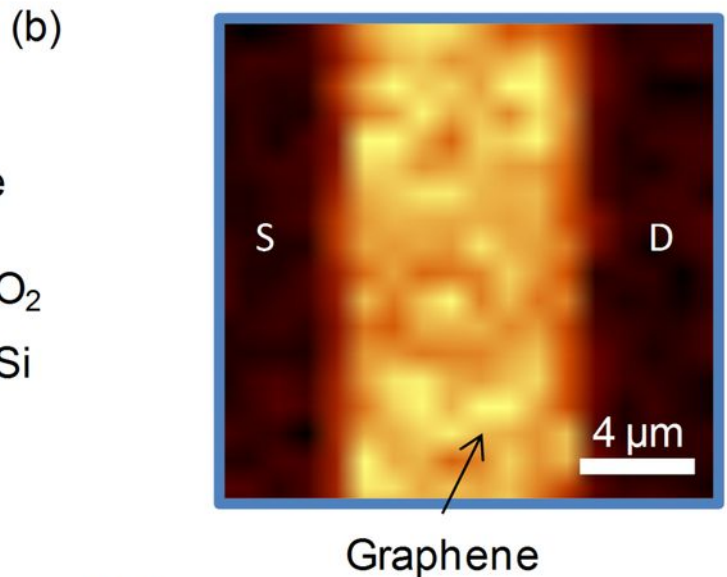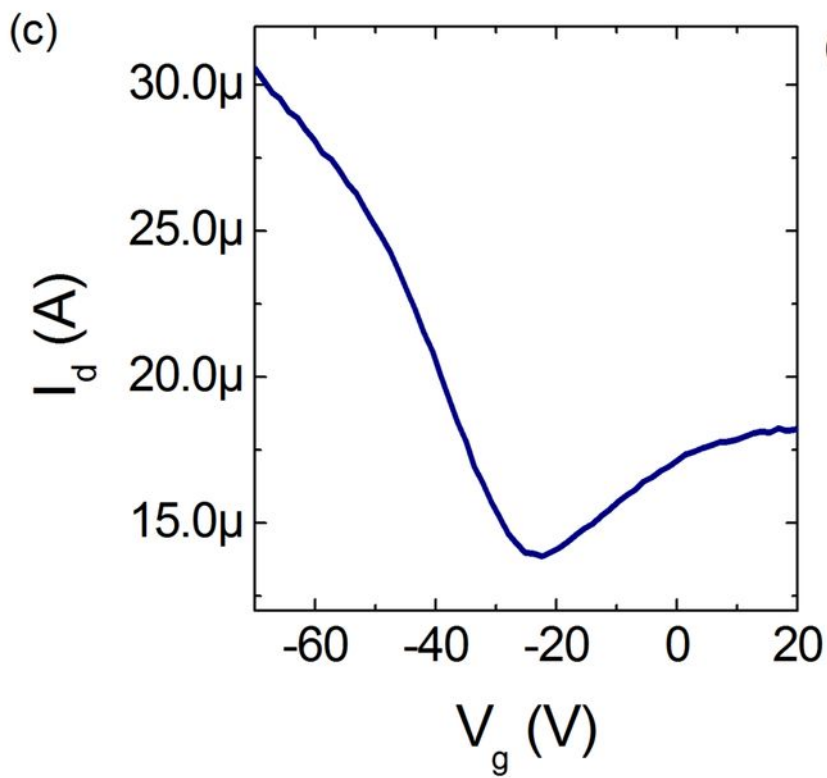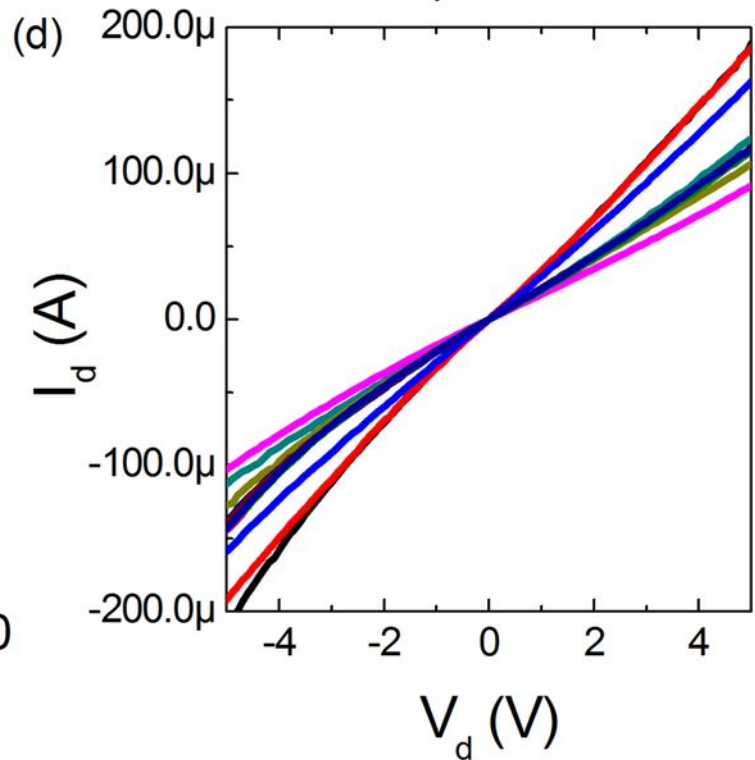